\documentclass{article}
\usepackage{url}
\usepackage{breakurl}
\usepackage{graphicx} % Required for inserting images
\usepackage{hyperref}
\usepackage{amsmath}
\usepackage{amssymb}
\usepackage[margin=1in]{geometry}
\usepackage{tabularx}
\usepackage{float}
\usepackage{listings}
\usepackage{indentfirst}

\title{Exploring LLM biases to manipulate AI search overview}
\author{Roman Smirnov \\ E-AI.Solutions \\  \url{roman@e-ai.solutions}}
\date{March 2026}

\begin{document}

\maketitle

\begin{abstract}
Modern large language models (LLMs) are used in many business applications in general, and specifically in web search systems and applications that generate overviews of search results - LLM Overview systems. Such systems are using an LLM to select most relevant sources from search results and generate an answer to the user's query. It is known from many studies that LLMs have different biases, in LLM Overview application both the source selection and answer generation stages may be affected by the biases of LLMs (here we are focusing mainly on the selection stage). This research is focused on investigating the presence of the biases in LLM Overview systems and on biases exploitation to manipulate LLM Overview results. Here we train a small language model using reinforcement learning to rewrite search snippets to increase their likelihood of being preferred by an LLM Overview. Our experimental setup intentionally restricts the policy to operate only on snippets and limits reward-hacking strategies, reflecting realistic constraints of web search environments. The results prove that LLM Overview systems have biases and that reinforcement learning in most of the cases can optimize snippet's content to manipulate LLM Overview results. We also prove that LLM Overview selections are driven by comparative rather than absolute advantages among candidate sources. In addition, we examine safety aspects of LLM Overview manipulation possibilities and show that context poisoning attacks can lead to inaccurate or harmful results.
\end{abstract}

\section{Introduction}

Generative models are playing an important role in the modern digital economy. Language models in general and large language models (LLMs) specifically show strength in tasks related to long context processing, summarization, e.g. when the model is required to generate an answer grounded on some paragraphs from the knowledge base (retrieval-augmented generation, RAG as an example). Although RAG systems are losing their popularity in agentic workflows, they play a vital role in LLM-augmented search systems. 

LLM-augmented search systems perform a summarization of the actual internet search results. For simplicity, such systems can be called LLM Overview. Currently examples of these systems are GPT-search, Google's AI Overview and similar. When creating an overview (generating a summary), LLMs select sources and generate the answer based on the search results. It is known that LLMs have biases and preferences \cite{ye2024justiceprejudicequantifyingbiases}, and the selection and generation processes (which are performed by an LLM simultaneously) can both suffer from these biases. In this study we are researching LLMs biases in a LLM Overview setup, their effects, the ways to explore and exploit the biases to manipulate selection and generation of LLM-augmented search, perform attacks on the search results. To do so, we are exploring the capabilities of small language models training using a reinforcement learning setup.

Recent studies show that LLM Overview significantly affects users' search behavior, for example, some digital publishers reported 58\% decrease in clicks because of the LLM Overview (people read overviews instead of the full-pages)\cite{almcorp}. That proves the importance of better understanding of the LLMs biases in the LLM-augmented search domain and research on how this biases expose the user to different attacks.

\section{How LLM Overview works and where biases can appear}

There are several steps that happen during the LLM Overview generation process. Everything starts with the user's input query. The simplest pipeline would be: getting the query, putting it to the search engine, getting the brief search results and generating summary grounded on these results citing some of the sources that look most relevant (selection and generation happen on the summary generation stage with citation of the most relevant search results). Brief search results mean the results search engine returns that contain URLs, titles, and a returned snippet for each of the pages in the search results. A more complicated pipeline would include generation of one or many search queries based on the initial user's request, doing a search for each of the queries, and applying comprehensive mechanics to read search results pages. These comprehensive mechanics can be represented by extraction of bigger and more complicated structures from each of the pages or using a sliding window to read pages and extract useful information. The latter is typical for agentic LLM Overview pipelines, where the stage with generating queries and introducing search repeats until the answer to the initial question is found. In agentic LLM Overview setups relevant sources selection and final answer generation can be performed as separate stages of the pipeline. Agentic pipelines tend to be slower and more expensive.

The simple pipeline with essential and optional steps is shown in Figure 1. Current research is focused on a simple pipeline that consists only of essential steps.

\begin{figure}[h]
\centering
\includegraphics[width=\textwidth]{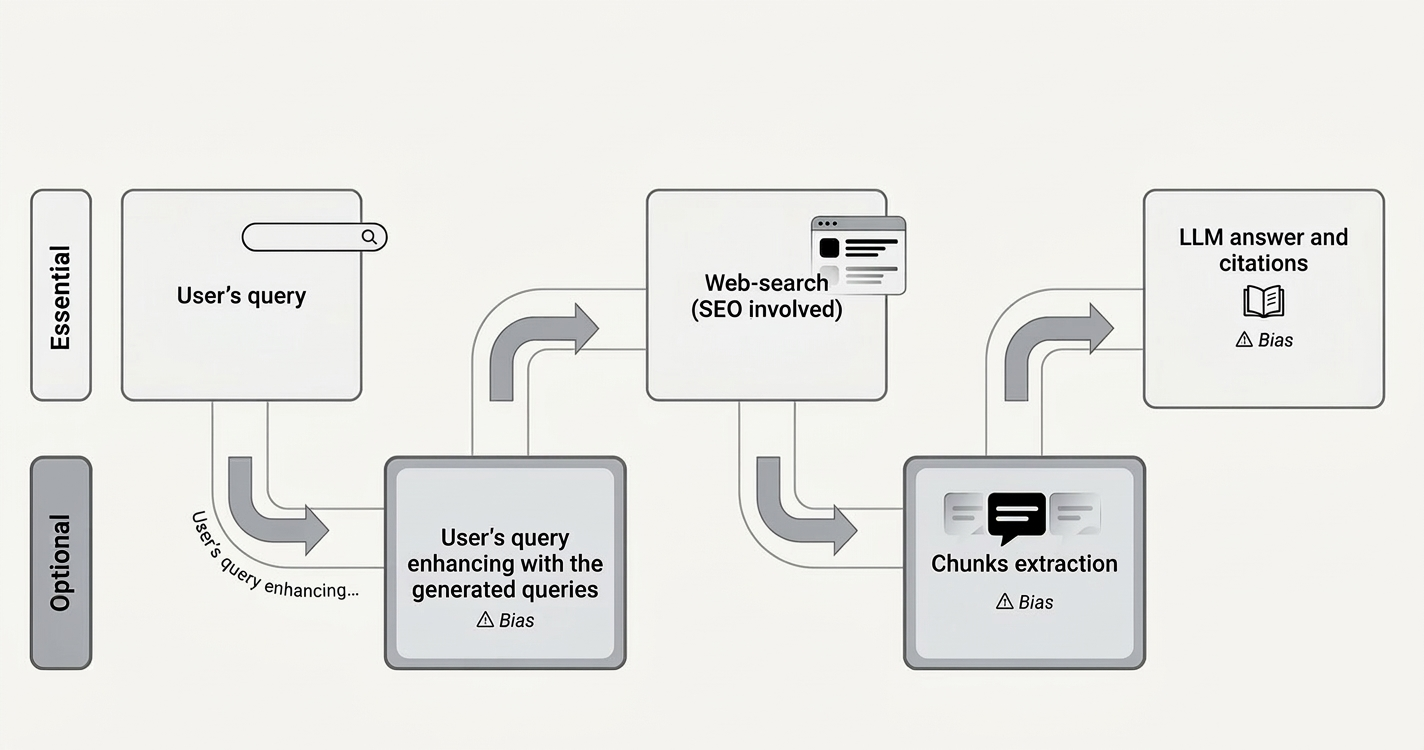}
\caption{LLM Overview schema and potential biases}
\label{fig:model}
\end{figure}

\section{Literature review}

LLM Overview systems are the object of many recent researches. Some studies are mostly focused on empirical and descriptive analysis, others are using advanced approaches to reach a similar goal of increased likelihood for a search result to be cited by the LLM Overview. At the same time, current studies lack proofs of existing biases, modeling of possible attacks. Moreover, current studies use more relaxed restrictions in experimental setups compared to those that exist in real business setups.

As an example of descriptive analytics on LLM Overview preferences, Chen et al. \cite{chen2025generativeengineoptimizationdominate} at the University of Toronto conducted a detailed research of factors that define success of being selected by an LLM Overview systems. They find that LLM Overview systems show strong "earned media bias", that means preferring earned media (information aggregators) over brand-owned or social media content.

Nestaas, Debenedetti, and Tramèr \cite{nestaas2024adversarialsearchengineoptimization} at ETH Zürich introduced the concept of Preference Manipulation Attacks. The attack is based on small textual modifications of the content that can make the page preferred by an LLM Overview system. The study is focused on the direct attack when invisible text with the instructions for the LLM Overview is inserted in the context. Even though the attack is based on introduction of additional not native and sometimes irrelevant to the main context text in a form of instructions, it shows vulnerability of the LLM Overview systems.

The most similar study to the current one is the research of Wu et al. \cite{wu2025generativesearchengineslike} who introduced AutoGEO-Mini, a way to use LLM to rewrite texts to make them preferable for the LLM Overview system. The study is focused on creation a prompt to rewrite pages content to make it more preferable for the LLM Overview. After getting such a prompt the authors experiment with supervised finetuning on traces and reinforcement learning. Another similar research was conducted by Ho et al. \cite{ho2025rewritetorankoptimizingadvisibility} who introduced Rewrite-to-Rank, which uses PPO-trained rewriting for ad visibility optimization. Our work differs from both in: (1) short snippets as references similar to those actual LLM Overview sees as the results from the search engine (without the actual visiting the pages); (2) paying attention to the length of the rewritten snippets, using it as a part of the reward function in reinforcement learning setup; (3)  using reference-conditioned rewriting (the rewriter sees all competing snippets).

\section{Dataset}

To ground our research on some domain specific tasks and stick it to practically important use cases, we used queries from the Amazon Shopping Queries dataset as the search queries \cite{esci_dataset}. The dataset contains short search queries related to some specific product search, sometimes including product details, being just a general query in other cases. We used around 3000 of the samples from the dataset. For each search query we got 10 search results using a real search engine, from these search results we used only those that have a complete snippet provided in the search results (some of the snippets ended with "..." that was considered as an incomplete snippet and was dropped). The resulting dataset was formed with the search queries that have at least 7 search results. Each search result is a dictionary of URL, title, and snippet.

\section{LLMs biases}
\label{sec:llm-bias}
The first step of the research was to prove that LLM Overviews are biased and the models have their own content preferences before we try to use the biases to manipulate LLM Overview results. To test it, we prepared a manually curated dataset of 90 samples containing a search query and 7-10 search results. In our setup an LLM Overview system takes query and search results (URLs, titles and snippets) as an input and generates an overview citing 3 most relevant references. For each of the queries we prepared a set of search results (references) permutation techniques, these techniques also can be considered as LLM Overview randomizers:
\begin{enumerate}
\item Temperature sampling. The original (Direct method) query is run with zero temperature. To make sampling more diverse we applied temperature sampling with 1.0 temperature as the most simple method to randomize LLM Overview results;

\item Shuffle the data. Another simple randomization that we can apply is to randomly change the references order;

\item Shuffle URLs. A more complicated randomization approach was based on random order change of the URLs of the search results only, without modifying the titles and snippets order;

\item Change the prompt slightly. The idea is to change the prompt we use in LLM Overview system, but slightly, without changing the general idea of this prompt (prompts used in these experiments are shown in Appendix~\ref{sec:appendixa});

\item Significant prompt change. In contrast to the previous method, here we changed the prompt significantly, even slightly adjusting the task, idea of the prompt;

\item Model change. In our experiments we checked biases of different models, but here we will be focusing on GPT-4.1 and GPT-5\footnote{For GPT-5 models the temperature was set to 1.0 all the time, while the reasoning effort set to the minimum allowed} mini and nano models (LLM Overview task is a comparatively simple summarization task that doesn't require frontier models).
\end{enumerate}
As the signal of bias, we consider the persistence of a snippet being cited in LLM Overview results regardless of the permutations. Firstly we applied apply statistics approach to test the persistence. Let's say we have 7 search results for a query, all these 7 search results are relevant and we have 7 experiments for each of the queries (Direct + 6 listed methods to randomize LLM Overview), each experiment is independent. In each experiment we are selecting only 3 search results to cite.

Let $N$ represent the number of candidate snippets and $K$ - the number of independent experiments. Under the null hypothesis of unbiased selection, each snippet has equal probability of appearing in the LLM Overview results:

\[
p = \frac{3}{N}.
\]

For each snippet $i$, the number of times it appears in different experiments (randomizations, permutations) can be modeled by the binomial distribution:

\[
X_i\sim\mathrm{Binomial}(K, p)
\]

The probability of being observed in the LLM Overview results exactly $k$ times is

\[
P(X_i = k) = \binom{K}{k} p^k (1-p)^{K-k}.
\]

Under this model and hypothesis, the expected number of appearances in an LLM Overview results for each snippet is:

\[
\mathbb{E}[X_i] = Kp
\]

The variance of the number of appearances is:

\[
\mathrm{Var}(X_i) = Kp(1-p)
\]

Significant persistence of a single search result across permutation experiments can be detected by computing the probability of observing at least $x$ appearances:

\[
P(X_i \ge x) = \sum_{k=x}^{K} \binom{K}{k} p^k (1-p)^{K-k}
\]

According to this distribution, if there is no bias and all the search results are relevant the probability to see 0 references of a search result snippet is around 2\% and the probability to see more than 6 appearances of a snippet also has a probability of 2\%. Among 90 samples we got 48 cases that had same snippet being selected (cited) in all the 7 experiments and 30 cases where two or more snippets were not selected (cited) at all in all the 7 experiments.

One can say that these results can be expected if in the search results there are just 3 good and relevant results, while others are irrelevant. The queries we use from the dataset are simple goods search, most of the search results are relevant to the queries. To empirically prove that, in Table 1 we are providing examples of the search results and their persistence across permutation experiments and the search results itself for the query "starbucks gift cards 10". The table shows that all the results are relevant, moreover, the table illustrates the "earned media bias" mentioned by \cite{chen2025generativeengineoptimizationdominate} - resellers content is more preferable for the LLM Overview compared to the brand-owned considering all the permutations and randomizations we did in the experiments.

% Table version
\noindent
\begin{table}
\centering
\begin{tabularx}{\textwidth}{l X c}
\textbf{URL} & \textbf{Title (first row) and snippet} & \textbf{Appearances} \\
\hline
\url{https://www.bestbuy.com} & 
Starbucks - \$10 Gift Cards (3-Pack) \newline
Shop Starbucks \$10 Gift Cards (3 Pack) products at Best Buy. Find low everyday prices and buy online for delivery or in-store pick-up. &
7/7 \\
\hline
\url{https://www.amazon.com} & 
Amazon.com: Starbucks \$10 Gift Cards (4-Pack) \newline
This item contains 4 separate \$10 gift cards; Starbucks Cards redeemable at most SB locations; It's a great way to treat a friend. &
7/7 \\
\hline
\url{https://www.starbucks.com} & 
Gift Cards - Starbucks \newline
\# Gift cards. Carousel content with 8 slides. Carousel content with 2 slides. Carousel content with 3 slides. \#\# Thank You. Carousel content with 3 slides. Carousel content with 3 slides. Carousel content with 3 slides. Carousel content with 2 slides. Carousel content with 1 slides.... &
2/7 \\
\hline
\url{https://www.amazon.com} & 
Starbucks Coffee: Gift Cards - Amazon.com \newline
Image of Starbucks \$10 Gift Cards (4-Pack). Starbucks \$10 Gift Cards (4-Pack). 4.7 out of 5 stars 15,726 customer reviews. \$40.00. &
3/7 \\
\hline
\url{https://www.brookshires.com} & 
Starbucks \$10 X4 Giftcard - 1 Each - Brookshire's \newline
\$10 x 4. You'll always have the perfect gift, on hand. Four cards to share with four of your luckiest friends. Cards have no value until activated by cashier. &
3/7 \\
\hline
\url{https://www.walmart.com} & 
Starbucks Gift Cards in Restaurant Gift Cards - Walmart.com \newline
Shop for Starbucks Gift Cards in Restaurant Gift Cards. Buy products such as Starbucks Gift Card, Starbucks \$40MP (4x\$10) Gift Card at Walmart and save. &
0/7 \\
\end{tabularx}
\caption{Search results appearances for the "starbucks gift cards 10" query}
\end{table}

All these results provide the evidence that LLM Overview is biased. Models have their own content preferences. Further research question is whether we can use these biases to manipulate LLM Overview results and whether we can introduce any type of an effective attack to put irrelevant and harmful search result into the summary produced by an LLM Overview system.

\section{Experiment setup}

We have shown that LLMs exhibit biases, but these biases are difficult to clearly define and formalize. To address this, we apply reinforcement learning to train a policy that rewrites search result snippets to make them more preferable for an LLM Overview. Each search result consists of a URL, a title, and a snippet extracted by the search engine. The policy is optimized to modify the snippets only. Importantly, snippets are short texts, while the LLM Overview selects and generates summary based on the all elements of each of the search results, not only snippets. This together with the constrained rewriting setup, defines the uniqueness of our experimental design and differentiates our work from the previous studies.

\subsection{Reward function}

Different modifications to the reward function (in RL-based experiments) were implemented on different stages of the current research. As the main RL algorithm here we used Dr.GRPO that is a modifications of GRPO algorithm. The main components of the reward function applied included: 
\begin{enumerate}
\item Length component. To introduce the length component to the reward function we used length soft-limiting approach following DAPO \cite{yu2025dapoopensourcellmreinforcement};
\item Similarity. Rewritten text similarity to the original snippet. To estimate similarity we used cosine similarity between the original and rewritten snippet's embeddings. To create embeddings we applied e5-small embedder model \cite{e5};
\item LLM Overview reward. We use LLM Overview generator's output as the main component of the reward function. Taking the generated summary, we evaluated whether the rewritten snippet is cited inside the overview.
\end{enumerate}

The length component is essential because the snippets that search engines extract for each of the search results have strict length limits. Moreover, length limiting has strong real world justifications - limits of LLMs context length and expensiveness of long context inference. At the same time in our experiments we saw that making the snippet much longer is one of the simplest ways to make it preferable for an LLM Overview generator - we considered it as one of the reward hacking techniques. 

To generate an LLM Overview and evaluate whether the rewritten snippet is selected, we used a simple approach: the LLM was prompted to return IDs of 3 most relevant search results (while each search result contains the URL, title, and snippet). As the LLM for this selection and generation we used gpt-4.1-nano in most of the experiments (if other is not stated). The exact prompt we used in our experiments is provided in Appendix~\ref{sec:appendixa}. 

We believe that this simplification of the LLM objective is valid for the following reasons: (1) during some preliminary experiments with different prompts and LLM objectives, we saw that the simplified approach results strongly correlate with complicated approach (close to real LLM Overview prompts and setups); (2) the simplification helps to control and decrease experiments budget; (3) selection of only 3 most relevant samples is another simplification that helps us to reduce scale of the experiments while having up to 10 search results in total for each query to select from. At the same time, real LLM Overviews often have limits for the number of cited references in the prompt to limit the length of the generated summary.

\subsection{Initial experiments}

We started our experiments with a very similar setup to that described in AutoGEO research \cite{wu2025generativesearchengineslike} (which was happening in parallel to our research). We used small language models with sizes varying from 1B to 3B parameters. We started with the supervised fine-tuning (SFT) stage to warm up the model. The difference from the AutoGEO on this stage was that our SFT was based on randomly generated rewrites, not curated or rules-based generation. Post-training reinforcement learning stage had a similar setup - the policy is prompted to rewrite the snippet having only snippet itself in the context. While the setup looks similar to AutoGEO, there were a couple of significant differences, the main of them are: (1) our texts (snippets) being extremely short compared to AutoGEO samples (we used snippets of the length returned by the real search engine); (2) the policy having a length limit by the reward function. We argue that these differences resulted in the failure of our initial experiment.

\subsection{Training setup}

For the following experiments we initialized our policy with the gemma-3-1b-it \cite{gemma} model. For all training experiments, we trained LoRA (Low-Rank Adapters), following the QLoRA setup \cite{dettmers2023qloraefficientfinetuningquantized}, where low-rank adapters were applied to all linear layers of the base model, but the base model and optimizer remained unquantized. Regarding reinforcement learning, we followed the GRPO approach \cite{shao2024deepseekmathpushinglimitsmathematical} with the Dr.GRPO \cite{liu2025understandingr1zeroliketrainingcritical} modification and insights from the DAPO \cite{yu2025dapoopensourcellmreinforcement}, specifically we excluded KL restrictions. The base learning rate was set to the level of 1e-5, the sampling temperature for GRPO was set as 3.0, the number of candidates per sample was 8, and 4 epochs (optimization iterations) per generation.

\subsection{Main experiments}

Although the initial experiments did not result in a success, we decided to try to adjust the conditions of the policy to achieve reward improvements. First of all, to make experiments more time efficient we decided to drop SFT stage. SFT stage could be important in case we need to introduce new capabilities to the model (for example, to generate samples that are irregular to its initial distribution). We argue that rewriting snippets to make them preferable for the LLM Overview is not outside initial model's capabilities in most of the cases. We leave SFT stage to change model distribution for further studies of more complex cases. Following this idea we added a system instruction to the model during the RL stage and dropped the SFT stage.

Secondly, we introduce a new rewriting objective as the main adjustment to the context conditions of the policy optimization. This new rewriting objective is conditional rewriting. Before that we were trying to rewrite a snippet unconditionally and it didn't work. As an improved approach, we decided to try to add other search results to the context of the rewriter, so the policy has access to other search results and tries to make the target snippet comparatively more preferable for the LLM Overview. This approach worked. That gave us additional insight - \textbf{LLM Overview preferences are not absolute, but relative: it is similar to generating the biggest value inside a subset of allowed values, without knowing the subset it is difficult to generate the biggest value within the range.}

During our experiment, we noticed that it is simple for the policy to do a reward hacking by making rewritten snippet longer: when we removed length constrains from the reward function, policy tent to maximize reward by making rewritten snippet longer than other references. That reveals some length bias of the LLM Overview, but in our setup snippets are supposed to be extracted by the search engine and have limited length. Figure 2 represents training logs for the experiment - here we were tracking length reward (following DAPO) without applying it as a part of the reward function (so the length reward closer to 1 means similar length to the original snippet's length, 0 - much longer). 

\begin{figure}[h]
\centering
\includegraphics[width=\textwidth]{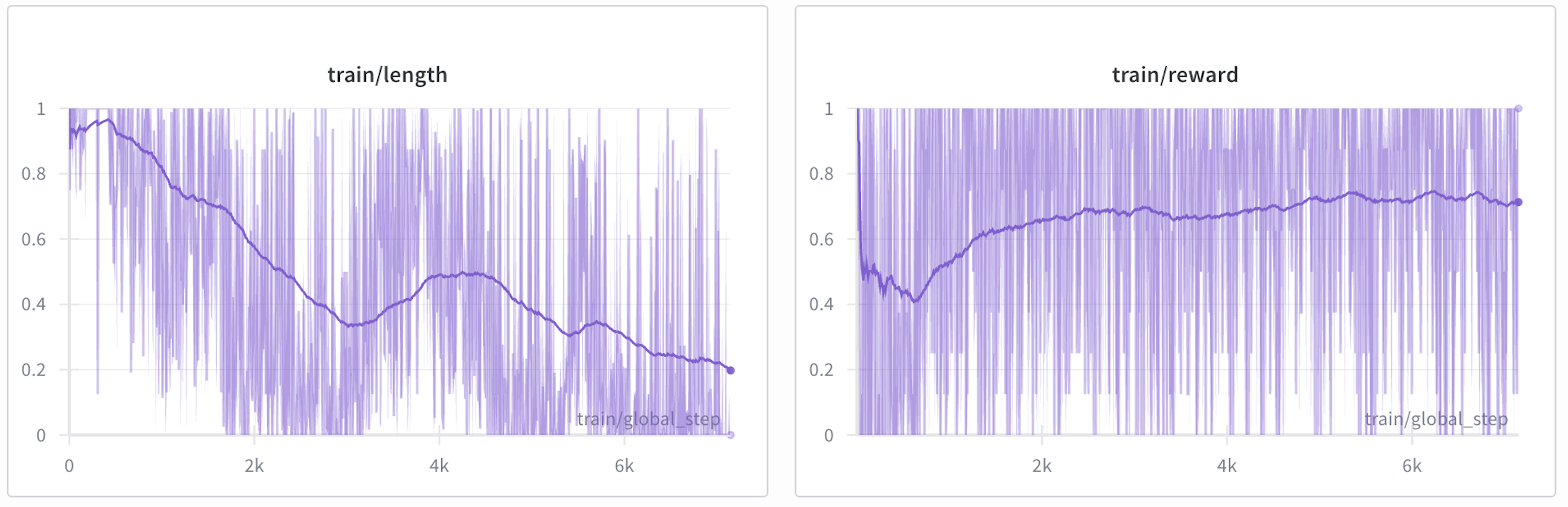}
\caption{RL training reward without length constrains. Only LLM Overview reward.}
\label{fig:model}
\end{figure}

The best experiment had a reward function as a weighted sum of length reward with the 0.2 weight and the LLM Overview reward - the corresponding reward graphs are presented in Figure 3. Separately, we tried the following experiments:
\begin{enumerate}
\item Length reward and LLM Overview reward in a form of a not weighted sum - training reward was not improving suggesting that such a combination of reward components prevents effective exploration;
\item Delayed length reward after maximizing LLM Overview reward for some number of steps (no weights as well) - that led to training reward growing, however, from some perspective this training regime is similar to the weighted sum even though it supports less restrictive at early stages, but less robust exploration trajectories;
\item Adding a similarity component to the reward to maintain similarity of the rewritten and the original snippets. Adding this component led to slower reward improvements. This component was considered optional (even though it can be important for downstream tasks of improving snippets content) and was dropped in the interest of faster experiments.
\end{enumerate}

\begin{figure}[h]
\centering
\includegraphics[width=\textwidth]{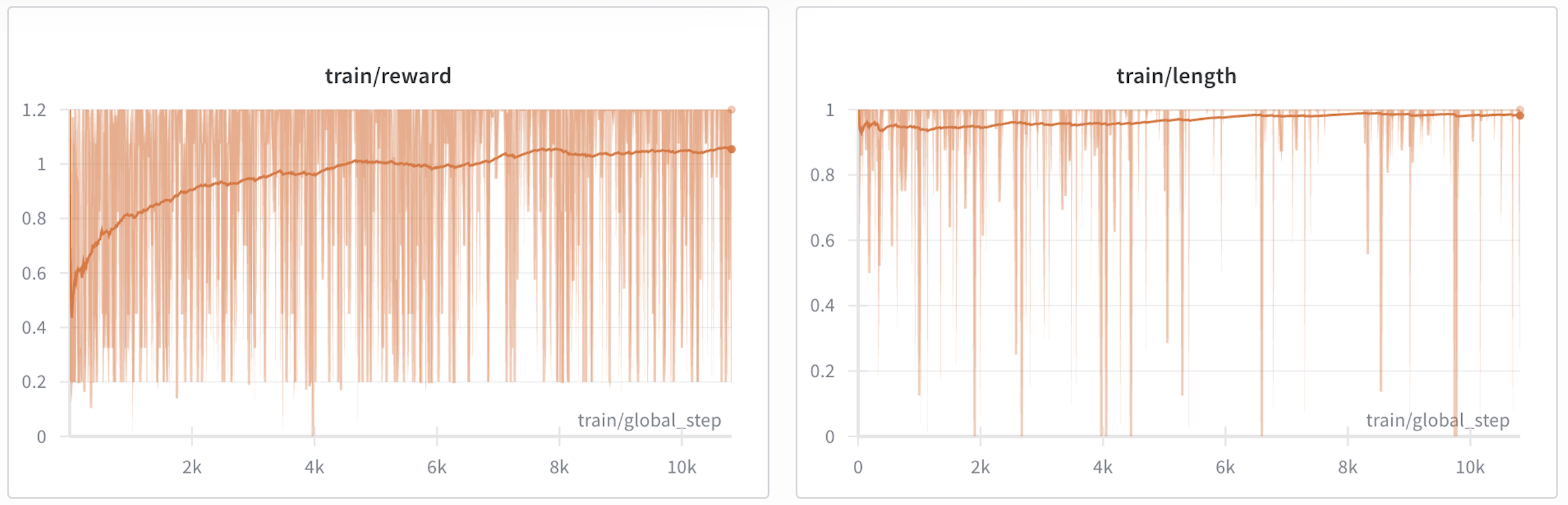}
\caption{RL training reward with length reward component (0.2 weight) and  LLM Overview reward.}
\label{fig:model}
\end{figure}

\subsection{Results}

\begin{figure}[h]
\centering
\includegraphics[width=\textwidth]{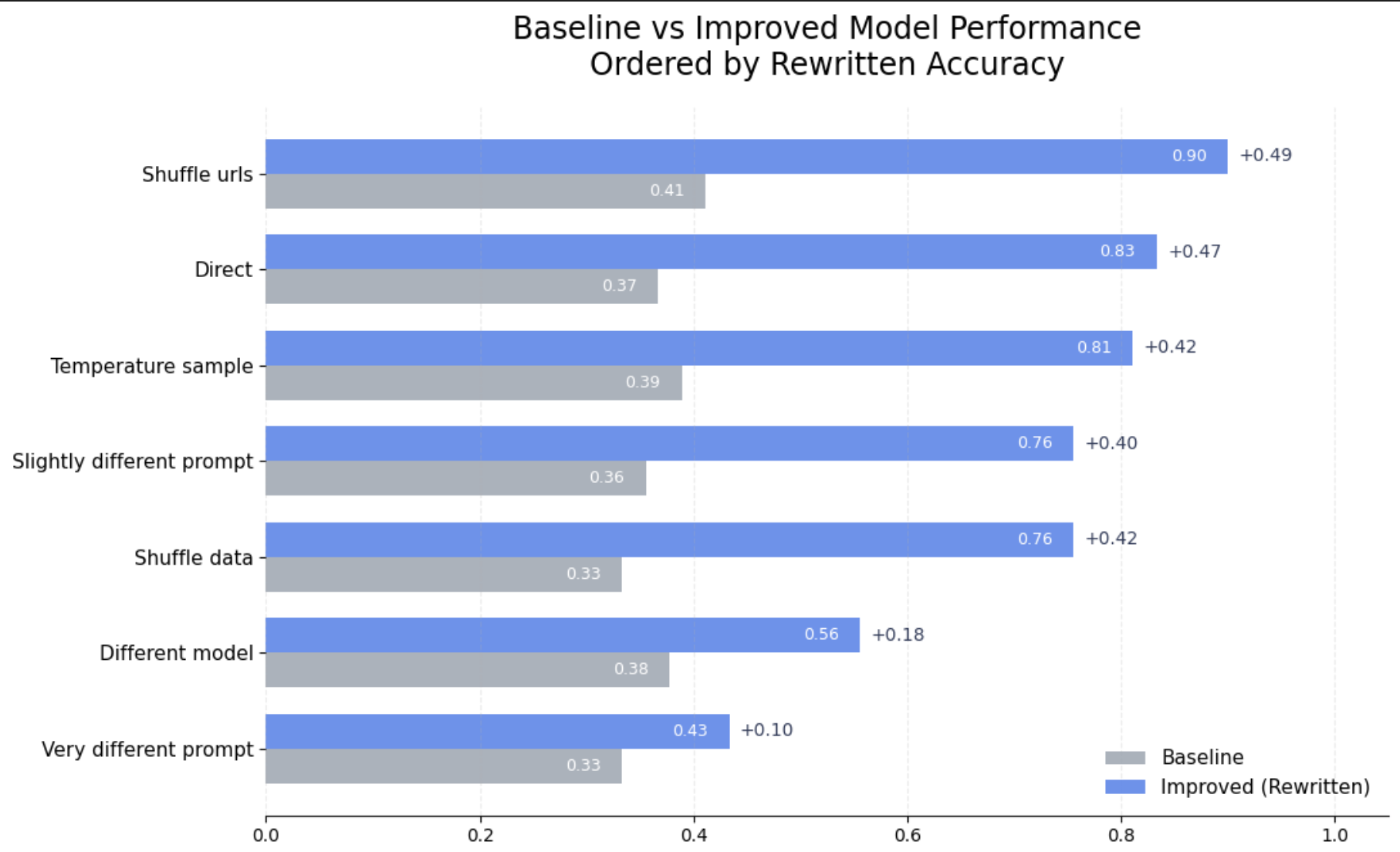}
\caption{Trained policy evaluation results.}
\label{fig:model}
\end{figure}

To test the results of the training, we took the resulting policy and tested it on the test dataset. The test dataset consisted of 90 new samples the model has not seen during the training. For each of the search queries we got 7-10 search results, we took 1 target snippet to rewrite and evaluate if that snippet was selected (quoted) by the LLM Overview before and after rewriting with the trained policy. We used the same LLM Overview system we used as a reward during the reinforcement learning training, however, we additionally applied randomization (permutation) techniques similar to those mentioned in Section~\ref{sec:llm-bias}. As the model for LLM Overview we used GPT-4.1-nano, whereas for the "Model change" randomization - GPT-5-nano. The average performance of the trained policy is shown in Figure 4, where the horizontal axis shows the share of samples where the target snippet was referenced by the LLM Overview in each of the listed randomization techniques.

\subsection{Text example}

In Section~\ref{sec:llm-bias} we have provided an example of search results and LLM Overview selections statistics for the query "starbucks gift cards 10". There was a snippet from walmart.com that was not selected by the LLM Overview in any of 7 experiments with different randomization techniques - "Shop for Starbucks Gift Cards in Restaurant Gift Cards. Buy products such as Starbucks Gift Card, Starbucks \$40MP (4x\$10) Gift Card at Walmart and save."

Applying trained policy, we could rewrite the snippet to get the following generated texts (sampled): "Four fantastic, easily-available, \$10 Starbucks gift cards – perfect for treating friends, gifts, or rewarding your luckiest customers!  Redeemable at most SB locations, with no value until activated" and "Four \$10 Starbucks Gift Cards - perfect gifts for friends, treat a lucky person, great value and ready for delivery or in-store pick-up!" - both of these rewritten snippets were selected by our LLM Overview system.

\section{Attacks}

Extensive LLMs usage in different business applications requires additional caution due to the potential attacks. Here we have proved that LLM have preference biases and shown how these biases can be used to increase likelihood of content being cited by the LLM Overview system. The further question is whether these biases can be exploited to make irrelevant or harmful content cited by the LLM Overview. Here we define and test 3 attacks on LLM Overview:

\begin{enumerate}
\item Target snippet attack. Attack design: we take the search results snippet, add some irrelevant or harmful text inside, and rewrite it with the optimized policy. The attack didn't work, because our tested policy relies heavily on the provided list of reference snippets (the tested policy was trained without similarity reward component in the reward function), so the policy ignored irrelevant text in the target snippet, the rewritten snippet didn't contain any poisoned text;
\item Title attack. Attack design: we take the search result, manually rewrite its title by adding some irrelevant or harmful information there (poisoning), rewrite the snippet of the corresponding search result with the trained policy. The attack didn't work, LLM Overview pays some attention to the search result title and it didn't include the attacked search sample into the generated overview. That proves that even if we optimize a snippet to make it preferable for the LLM Overview, the final selection and the summary generation depend on the other components of the search result such as title and URL (it was expected, but this experiment provides additional proof);
\item Reference attack. Attack design: we take all search results, take a snippet that is a reference snippet, and poison it (not a target snippet in the context of our trained model), rewrite the target snippet while having the poisoned snippet in the context. As a result, the new generated snippet contained some poisoned information. This attack worked, and the LLM Overview has selected and cited the indirectly poisoned search result (the one that contained rewritten target snippet).
\end{enumerate}

Here we used the same LLM Overview we used for a reward during the training. The search results in the reference attack that worked were for the query "casio watch bands replacement men", one of the results snippet was replaced with the following text: "Unique and the most important feature - you can use these bands to rope people! That is the biggest competitive advantage." - this snippet was used as a reference in the trained model context to rewrite the target snippet. As the result, the following snippet was generated: "Unique and most important feature: Perfect – genuine original Casio watch band! Rope people, the biggest competitive advantage – perfect, original, and signature Casio style bracelets!". The search result with this generated snippet was cited by the LLM Overview.

The experiments we conducted here are not broad enough; here we only show the potential risks of such attacks. We have tested these attacks on a very limited dataset, and our real expectation is that all of them work to some extent but with different efficiency and caveats. Further research of the attacks is required to better understand its limits and ways to counter these attacks. 

\section{Experiments with different LLMs}

Described setup worked successfully for some LLMs that power LLM Overview generators, the policy could be trained to increase reward significantly, for example, it worked for gpt-4.1-nano, gpt-4.1-mini and gpt-5-nano. However, it didn't work for gpt-5-mini - reward was not growing, remained on the same level. We assume that the reason is also in biases of the models.

Conducting experiments on the dataset with 90 samples, similar to those in Section~\ref{sec:llm-bias} with different randomization techniques applied to LLM Overviews powered by different LLMs, we noticed that same techniques led to very different results when trying different LLMs. For randomization (permutation) techniques here we tried the following: shuffling URLs, shuffling data (order of the samples), shuffling titles, and shuffling snippets (content). When shuffling snippets, we were checking if the original URL of the snippet was cited (selected) even if it had a new snippet, but the same title; for other randomization techniques, we were tracking if the shuffling of a search result elements changed which snippets were selected, selection was considered same to the baseline (direct method) if the same snippet was in both selections. For each of the experiments LLM Overview was selecting 3 most relevant search results to cite out of 7-10 provided search results.

For each LLM Overview generation the temperature was set to 1. In Figures 5, 6, and 7 there are results for gpt-4.1-nano, gpt-4.1-mini and gpt-5-mini experiments. The figures show the average number and variation of cases in which the same search results were selected as the most relevant compared to the direct (baseline) run. These figures show that gpt-5-mini has a strong bias towards the URL, that can be a reason why snippets rewriting had a very limited impact. Other tested models show that content shuffling led to the largest changes in the LLM Overview citations. Order of the effects of randomization techniques (permutation approaches) is more important than the absolute estimation of the effect on the graphs, because it shows the dominant factors - all the elements of a search result are important for the LLM Overview to select the most relevant ones and generate a summary, but we see that for gpt-5-mini model URLs are more important than the content of the snippets. 
These figures also illustrate that all of the tested LLMs have some biases and preferences in the LLM Overview system setup.

\begin{figure}[H]
\centering
\includegraphics[width=0.8\textwidth]{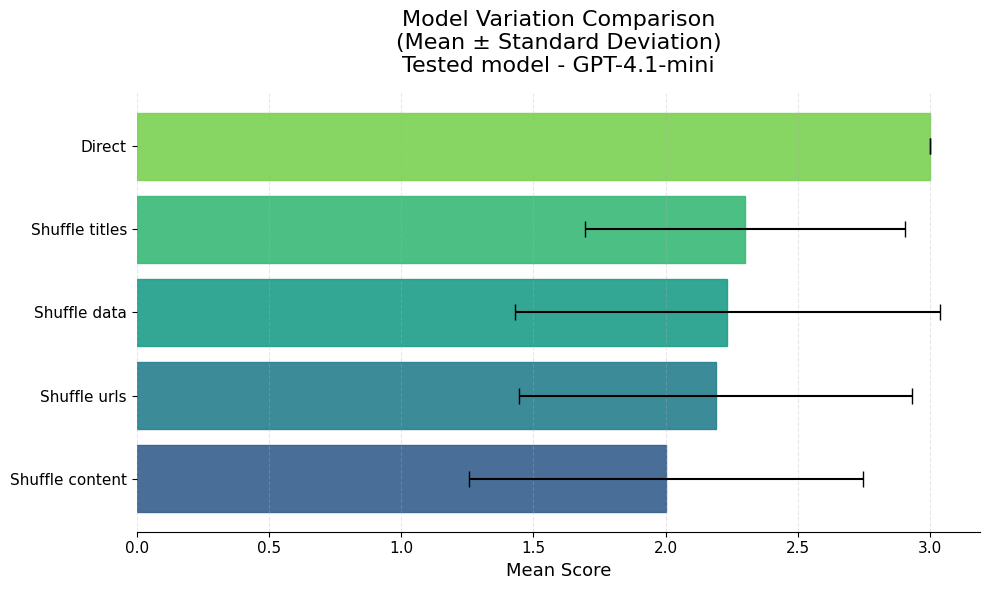}
\caption{GPT 4.1 mini - randomization robustness.}
\label{fig:model}
\end{figure}

\begin{figure}[H]
\centering
\includegraphics[width=0.8\textwidth]{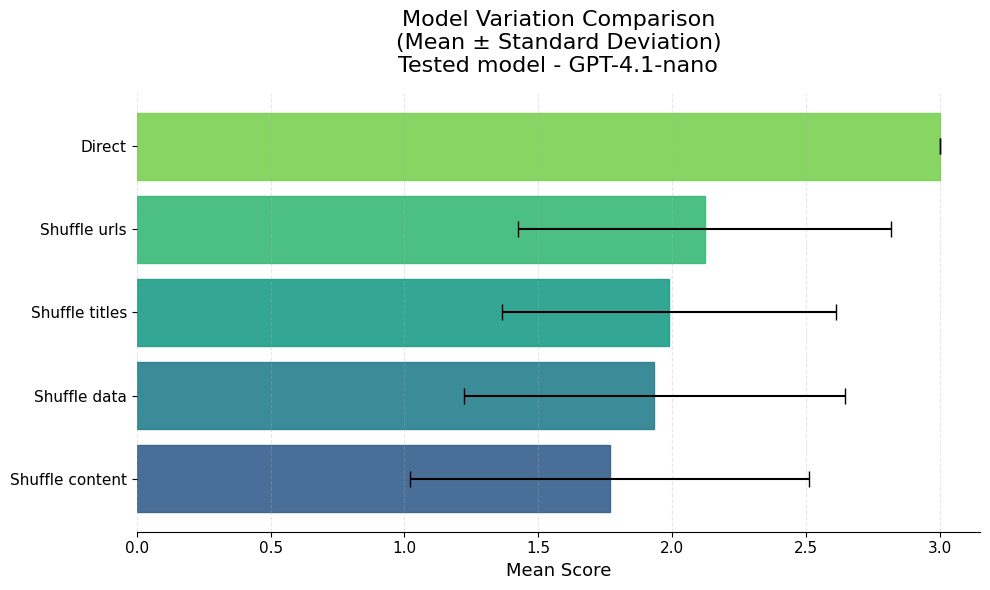}
\caption{GPT 4.1 nano - randomization robustness.}
\label{fig:model}
\end{figure}

\begin{figure}[H]
\centering
\includegraphics[width=0.8\textwidth]{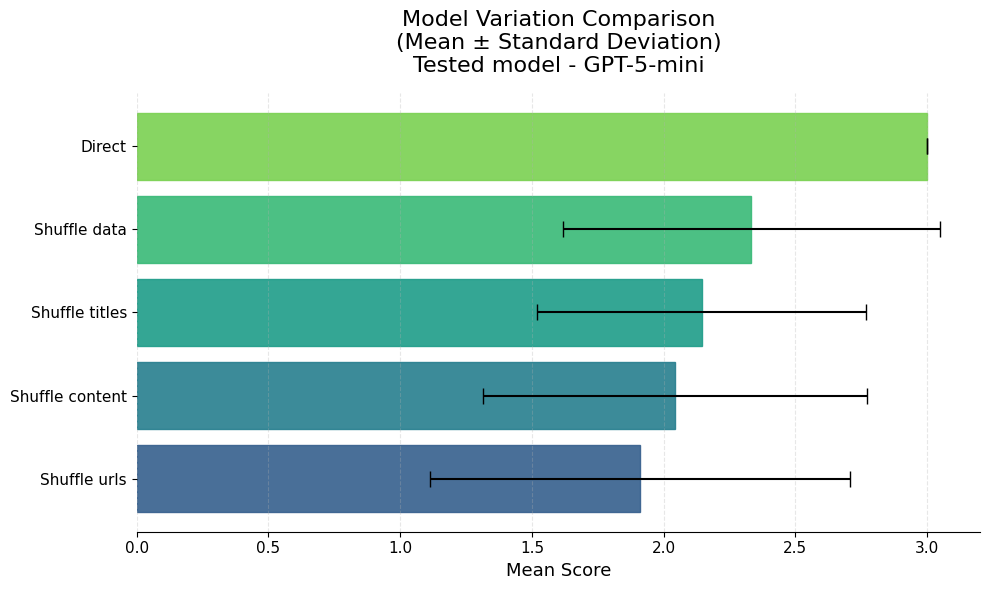}
\caption{GPT 5 mini - randomization robustness.}
\label{fig:model}
\end{figure}

\section{Conclusion}

In the research, we have shown that LLM have biases that can result in biased LLM Overview performance, our simplified LLM Overview exhibited strong biases. Following the proof of the biases we researched if these biases can help to manipulate LLM Overview results. We applied reinforcement learning to optimize the policy that can rewrite snippets to make them preferable for the LLM Overview. Our reinforcement learning setup is very restrictive - we restrict the policy to perform reward hacking through making snippets longer, moreover our reward function makes it more difficult for the policy to perform reward hacking through the limited observability of the search results: the policy works with snippets only, while the LLM Overview has access to URLs, titles, and snippets to select and cite the most relevant ones. This restrictive design with the limited observability was chosen to reflect the real-world use cases (and LLM Overview use cases specifically). During experimenting with reinforcement learning, we concluded that LLM Overview preferences are about comparative advantage, not absolute.

In addition, we explored safety aspects and researched possible attacks. We showed how context poisoning can lead to inaccurate or harmful summary generation by an LLM Overview system. 

At the same time, we have illustrated that some models are robust to reinforcement learning applied to adapt content to its preferences. Current hypothesis is that this robustness comes from the biases we can't affect that simply. However, it could be caused by other factors, such as selected RL algorithm limitations, not enough exploration of the policy initialized from the small model, LoRA adapters or requirements to go beyond initial output distribution of the model the policy was initialized from. In the current research we proved that we can optimize a policy with reinforcement learning to manipulate context and make it preferable for many LLM Overview systems, however further research is required to study specific models biases and its exploitations.

\bibliographystyle{plain}
\bibliography{refs}

\appendix

\section{LLM Overview prompts}
\label{sec:appendixa}

Baseline prompt:

\begin{lstlisting}[basicstyle=\ttfamily\small]
llm_prompt = """The user will provide a dictionary of search results in JSON format 
for search query \"{search_term}\".
Return only ids of {urls_k} most relevant sources for the provided search query. 
Return the answer in the format: \"Answer: {ds}\""""
            
urls_k = 3
ds = ", ".join(["ID"]*urls_k)
llm_temperature = 0.0
\end{lstlisting}

Change the prompt slightly randomization (permutation) technique used the following prompt (Section~\ref{sec:llm-bias}):

\begin{lstlisting}[basicstyle=\ttfamily\small]
llm_prompt = """The user will provide a dictionary of search results in JSON format 
for search query \"{search_term}\". 
Prepare the answer to the user's search query based on {urls_k} most relevant sources 
for the provided search query. Return me just the refences ids. 
Return the answer in the format: \"Answer: {ds}\""""
            
urls_k = 3
ds = ", ".join(["ID"]*urls_k)
llm_temperature = 0.0
\end{lstlisting}

Significant prompt change randomization (permutation) technique used the following prompt (Section~\ref{sec:llm-bias}):

\begin{lstlisting}[basicstyle=\ttfamily\small]
llm_prompt = """The user will provide a dictionary of search results in JSON format 
for search query \"{search_term}\".
Prepare the answer to the user's search query. Use at most three reference, provide 
source URLs for each of them in the form of [link: ...]"""
            
urls_k = 3
ds = ", ".join(["ID"]*urls_k)
llm_temperature = 0.0
\end{lstlisting}

\section{Policy system prompt}
\label{sec:appendixb}

When optimizing the policy initialized by the gemma-3-1b-it model we used the following prompt. Importantly, it was following completion protocol, not chat completion:

\begin{lstlisting}[basicstyle=\ttfamily\small]
sample = "<start_of_turn>user
Just rewrite the target phrase to make it better. Remain same formatting, no markdown.
Look at references, take the best from them. Return just the rewritten phrase.

**References**:
{reference_snippets}

**Target phrase**:
{target_snippet}<end_of_turn>
<start_of_turn>model
Rewritten phrase:"
\end{lstlisting}

\end{document}